\documentclass[%
reprint,
amsmath,amssymb,
aps,
pra,
]{revtex4-2}

\usepackage{graphicx}
\usepackage{dcolumn}
\usepackage{bm}
\usepackage{hyperref}
\usepackage{float}
\usepackage[table]{xcolor}
\usepackage{array}
\usepackage{makecell}
\usepackage{colortbl}
\usepackage{siunitx}
\usepackage[version=4]{mhchem}

\DeclareSIUnit{\dBm}{dBm}

\definecolor{lightgray}{gray}{0.55} 

\newcolumntype{A}{>{\hspace{8pt}\cellcolor{lightgray}}c<{\hspace{8pt}}}

\begin{document}
\title{DC-powered broadband quantum-limited microwave amplifier}
\author{N. Nehra}
\author{N. Bourlet}
\author{A. H. Esmaeili}
\author{B. Monge}
\author{F. Cyrenne-Bergeron}
\author{A. Paquette}
\author{M. Arabmohammadi}
\author{A. Rogalle}
\author{Y. Lapointe}
\author{M. Hofheinz}
\affiliation{Institut Quantique, Université de Sherbrooke, Sherbrooke, Quebec J1K 2R1, Canada}

\date{\today}

\begin{abstract}
Fast, high-fidelity, single-shot readout of superconducting qubits in quantum processors demands quantum-limited amplifiers to preserve the optimal signal-to-noise ratio. Typically, quantum-limited amplification is achieved with parametric down-conversion of a strong pump tone, which imposes significant hardware overhead and severely limits scalability. Here, we demonstrate the first DC-powered broadband amplifier operating within 0.2 photons of the quantum limit. Our impedance-engineered Inelastic Cooper-pair Tunneling Amplifier (ICTA)—a voltage-biased SQUID in which Cooper pairs tunnel inelastically by emitting signal–idler photon pairs—operates in reflection, delivering \SI{13}{\decibel} of average gain across a \SI{3.5}{\giga\hertz} bandwidth in a single stage. Semiclassical simulations accurately predict the gain and saturation power, enabling further design improvements. By eliminating the pump-tone infrastructure, the broadband ICTA promises to dramatically reduce the hardware complexity of quantum-limited amplification in superconducting quantum processors.
\end{abstract}

\maketitle

Superconducting qubits are typically read out by dispersively coupling them to a readout resonator. The state-dependent frequency shift is then measured via the phase shift of a probe tone sent to the resonator. However, during this measurement, the intracavity photon number must remain low to avoid measurement-induced state transitions~\cite{DSank2016MIST,MKhezri2023MIST}. Discriminating the required low-amplitude signals with sufficient fidelity requires quantum-limited amplifiers, such as Josephson parametric amplifiers (JPAs)\cite{CBeltran2008JPA, Bergeal2010JPA}, adding the minimum noise allowed by quantum mechanics and offering the highest readout fidelity.

Early Josephson parametric amplifiers (JPAs) suffered from narrow bandwidth and low saturation power. Substantial improvements in both figures of merit have been demonstrated through advanced designs~\cite{MutusWhite2014WideJPA,TRoy2015BeyondGBWP,ONaman2019HighSatJPA,TWhite2023HighSatJPA}. Josephson traveling-wave parametric amplifiers (JTWPAs) eliminate these trade-offs by distributing the nonlinearity along a transmission line, routinely achieving gain-bandwidth products over $\SI{10}{\giga \hertz}$ and saturation powers above $\SI{-100}{dBm}$ at the price of requiring much stronger pump tones in the $\SIrange{-85}{-65}{\dBm}$ range~\cite{Yaakobi2013TWPA,Macklin2015RPM,Ranadive2022KerrRev}. This breakthrough has enabled high-fidelity multiplexed qubit readout~\cite{EJeffery2014FastReadout, Heinsoo2018MultiQreadout, Peter2025FastMultiplexed} and high-power qubit readout using nonlinear coupling~\cite{mori2025highPowReadout}.

Despite these advances, both JPAs and JTWPAs demand intensive hardware, including complex microwave circuitry for generating, routing, and filtering the required pump tone, hindering scaling of superconducting quantum processors. JTWPAs further require advanced microwave engineering and precise fabrication control, complicating scalable readout architectures.

In contrast, DC-powered amplification eliminates the challenges of managing strong pump tones, significantly simplifying quantum-limited amplification. However, DC-powered amplifiers based on resistively shunted Josephson junctions (JJs) have long struggled to achieve quantum-limited noise performance~\cite{DHover2012SLUG, Lahteenmaki2012SJA}. More recently, Inelastic-Cooper-pair-Tunneling Amplifiers (ICTAs), which use voltage-biased unshunted Josephson junctions, have achieved near quantum-limited noise, but with limited gain-bandwidth products ranging from $\SIrange{0.1}{0.65}{\giga\hertz}$ \cite{Jebari2018ICTA, UMartel2025VBinfluence}. These amplifiers operate similarly to a three-wave mixing parametric amplifier, in which the pump is replaced by inelastic Cooper-pair tunneling: the Josephson junction is biased at a voltage $V_{\text{dc}}$ below the superconducting gap, such that the energy $2eV_{\text{dc}}$ released by a Cooper pair, tunneling inelastically through the junction, corresponds to the sum of photon energies of the signal and idler modes of the amplifier.

So far, achieving broadband amplification with these devices has proven difficult because photon emission in inelastic Cooper-pair tunneling scales, to first order, with $\operatorname{Re}\, Z_{\text{JJ}}(f)/f$, where $Z_{\text{JJ}}$ denotes the input impedance seen by the Josephson junction from the embedding linear circuit~\cite{GIngold1992PE,MHofheinz2011BrightSide}. This scaling strongly favors parametric amplification—and ultimately oscillations—of out-of-band signal–idler pairs close to DC and $2eV_{\mathrm{dc}}/h$. Only when the desired signal–idler pair has high characteristic impedance (relative to the bias-circuit impedance) and high quality factors can sufficient gain be achieved before the amplifier becomes unstable. Recent progress in voltage-bias circuits with low, flat output impedance~\cite{UMartel2025VBinfluence} allows for lower-$Q$ signal–idler modes while preserving stability. This advance enables bandwidth-engineering techniques—originally developed for JPAs~\cite{MutusWhite2014WideJPA,TRoy2015BeyondGBWP,ONaman2022Synthesis}—to be directly transferred to ICTAs.


\begin{table*}[htbp]
\centering
\renewcommand\arraystretch{1.3}
\arrayrulecolor{black}
\caption{Final targeted values of circuit components in Fig.~\ref{fig:fig1}(b)}
\definecolor{lightgray}{gray}{0.9} 
\label{tab:tab1}
\begin{tabular}{!{\vrule width 1.5pt} A !{\vrule width 1.5pt} A !{\vrule width 1.5pt} A !{\vrule width 1.5pt} A !{\vrule width 1.5pt}}
\Xhline{2\arrayrulewidth} 
Parallel ($L_{p},\hspace{0.1 cm} C_{p}$) [orange]  & Series ($L_{s},\hspace{0.1 cm} C_{s}$) [blue] & $\lambda/4$ resonator ($Z_0, f_0\hspace{0.1 cm}$) [brown] & $\text{Max } I_c$ (per junction) [yellow]
\\
\Xhline{2\arrayrulewidth} 
1.38 nH, 530 fF   & 1.94 nH, 373 fF   & $58.8\hspace{0.2 cm} \Omega$, 5.88 GHz & 600 nA  \\
\Xhline{2\arrayrulewidth} 
\end{tabular}
\end{table*}

\begin{figure}[htbp]
\centering
\includegraphics[%
width = 0.45\textwidth]{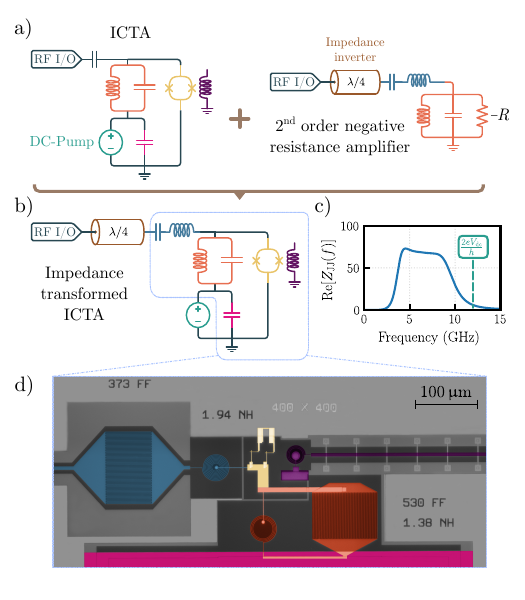}
\caption{\label{fig:fig1} \textbf{Impedance-Transformed ICTA.} \textbf{(a)} Left: Single-resonator ICTA, featuring a lumped-element LC resonator (orange) in series with a SQUID loop (yellow). A DC source (turquoise) biases the SQUID through the inductor of the $LC$ resonator, while a large capacitor (magenta) shunts the resonator to ground at RF frequencies. An on-chip flux bias line (purple) enables tuning of the SQUID’s critical current. Right: An impedance transformed negative resistance amplifier (NRA) based on a second-order resonator-network. \textbf{(b)} Circuit diagram of an impedance-engineered ICTA combining single-Resonator ICTA with a second order network employing a lumped-element series $LC$ resonator paired with a $\lambda/4$ coplanar waveguide (CPW) resonator to transform the ICTA's input impedance for broadband amplification. \textbf{(c)} Circuit impedance seen by the Josephson junctions of the SQUID for the optimized design parameters. \textbf{(d)} Zoomed-in grayscale optical microscope image of the device, excluding the $\lambda/4$ CPW resonator of the impedance inverter. All elements are color-coded as in the circuit diagram, light gray denotes two niobium layers (ground plane), darker gray denotes a silicon nitride layer atop a single niobium base, and black denotes the absence of niobium layers.
}
\end{figure}

We design bandwidth-engineered ICTAs (Fig.~\ref{fig:fig1}(a)) starting from a compact lumped-element core: a parallel LC resonator that hosts both the signal and idler modes is connected to a superconducting quantum interference device (SQUID), which is DC voltage biased through the resonator. A large on-chip capacitor ($\SI{100}{pF}$) on the voltage-bias side ensures an RF ground, while an on-chip flux bias line threads magnetic flux through the SQUID loop to modulate its critical current. This setup establishes a robust voltage-biasing scheme and enables the application of impedance transformation techniques inspired by negative resistance amplifiers (NRAs)\citep{ONaman2022Synthesis}. To achieve the desired impedance characteristics, a lumped-element series LC resonator is employed to introduce a positive slope in the reactance seen by the ICTA. An impedance inverter, implemented by a quarter-wavelength (\(\lambda/4\)) coplanar waveguide (CPW) resonator, accounts for the impedance mismatch between the network impedance and the standard $\SI{50}{\ohm}$ impedance (Fig.~\ref{fig:fig1}(b)). Initial component values were derived from second-order Butterworth negative-resistance amplifier prototype coefficients \cite{ONaman2022Synthesis}, targeting $\SI{20}{dB}$ gain, a network impedance of $\SI{81.7}{\ohm}$, and $\SI{25}{\percent}$ fractional bandwidth ($\Delta f /f$) centered at $\SI{6}{GHz}$. These prototype values directly informed the choice of lumped-element capacitances and inductances for the series and parallel LC resonators, as well as the characteristic impedance of the $\lambda/4$ impedance inverter. Minor adjustments were made to the lumped-element values to compensate for parasitic junction capacitance and biasing-network effects. The resulting final component values for the impedance-transformed ICTA are listed in Table~\ref{tab:tab1}. We used these values to calculate the effective embedding impedance presented to the SQUID loop. The design was first validated using a linearized coupled-mode analysis of the parametric coupling between the signal and idler modes. The parameters were subsequently fed into our in-house semiclassical gain simulator, which incorporates all higher-order mixing processes (see Supplementary~\ref{sec:simulations}). The inclusion of all mixing terms strongly degrades numerical convergence at high gain; therefore, the presented simulations are restricted to moderate gain regimes where fast and stable convergence is achieved.



We analyze the gain of the ICTA as a function of two key control parameters: the DC voltage bias across the junctions, expressed as Josephson frequency $  f_{\text{dc}} = 2eV_{\text{dc}}/h $ , and the effective critical current $  I_{\text{c}}  $ of the SQUID, controlled by the applied magnetic flux.
The Josephson frequency directly corresponds to the pump frequency in conventional 3-wave mixing JPAs and the critical current of the SQUID to the pump amplitude.

\begin{figure}[htbp]
\centering
\includegraphics[%
width = 0.45\textwidth]{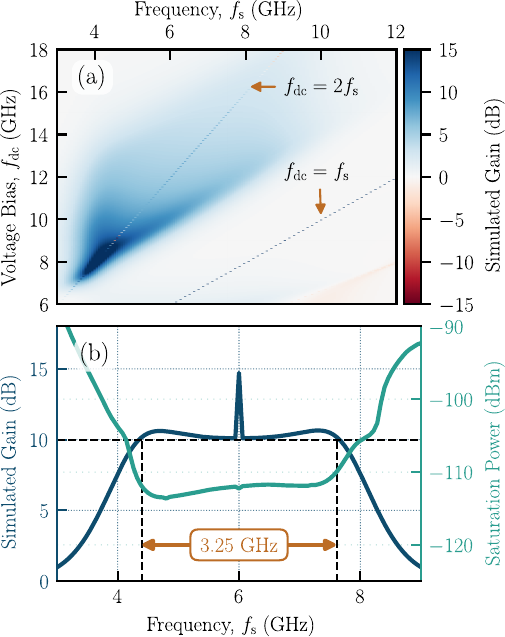}
\caption{\label{fig:fig2} \textbf{Semiclassical simulations of impedance transformed ICTA} \textbf{(a)} Gain characteristics as a function of signal frequency, $f_{\text{s}}$, and voltage bias expressed as Josephson frequency $f_{\text{dc}} = \frac{2e}{h}V$ with $\frac{2e}{h} \approx \SI{0.5}{\giga\hertz/\micro\volt}$, at $I_{\text{c}} =\SI{200}{\nano\ampere}$. High gain (blue) is achieved in a parallelogram-shaped region where the circuit impedance seen by the SQUID (see Fig.~\ref{fig:fig1}(c)) is high at both signal frequency $f_\text{s}$ and idler frequency $f_\text{i} = f_\text{dc} - f_\text{s}$. The sharp lines are due to the AC Josephson effect (slope 1) and degenerate parametric amplification (slope 2).  \textbf{(b)} Simulated gain and input-referred \SI{1}{\decibel} compression point of the ICTA biased at ($f_{\text{dc}} = \SI{12}{\giga\hertz}$, $I_{\text{c}} = \SI{280}{\nano\ampere}$), as a function of signal frequency.
}
\end{figure}

To ensure simulation convergence, we first sweep $f_{\text{dc}}$ over a wide range while keeping $I_{\text{c}}$ low and the input signal power small ($\SI{-140}{dBm}$). This yields a 2D gain map with signal frequency $f_{\text{s}}$ on the x-axis and $f_{\text{dc}}$ on the y-axis (Fig.~\ref{fig:fig2}(a)). The map shows a region of high gain forming a parallelogram, where signal frequency $f_{\text{s}}$ and idler frequency $f_{\text{i}} = f_{\text{dc}}-f_{\text{s}}$ lie within the designed high-impedance environment (see Fig.~\ref{fig:fig1}(c)). The upper boundaries of the parallelogram are blurred due to slower roll-off of $\text{Re}[Z_{\text{JJ}}(f)]$, at the upper edge of the band (Fig.~\ref{fig:fig1}(c)). The sharp lines appearing in the plot correspond to the ac Josephson effect at $f_{\text{dc}} = f_{\text{s}}$ and degenerate amplification at $f_{\text{dc}} = 2 f_{\text{s}}$. 

We select the operating point ($f_{\text{dc}} = \SI{12}{\giga\hertz}$, $I_c = \SI{280}{\nano\ampere}$) to center the amplification band at the target frequency while keeping $I_c$ just below the value at which the full nonlinear simulations fail to converge. This critical current is below the initial target and delivers the gain profile in Fig.~\ref{fig:fig2}(b) with lower gain but higher bandwidth: The simulations predict a nearly flat \SI{10}{\decibel} gain across a \SI{3.25}{\giga\hertz} bandwidth with an average input-referred \SI{1}{dB} compression point of \SI{-113}{\dBm}.


These simulations inform the final layout of the optimized amplifier design, which undergoes finite element electromagnetic simulations in COMSOL Multiphysics to accurately model the inductors and capacitors of the circuit. The device is fabricated using a process featuring two niobium routing layers separated by silicon-nitride dielectric and aluminum-based Manhattan-style Josephson junctions (\hyperref[sec:fabrication]{Supplementary}). Following fabrication, the device is packaged in a copper box, mounted to the baseplate of a dilution refrigerator, and wired to enable precise gain and noise characterization at millikelvin temperatures. The amplifier is protected from stray magnetic fields with a room-temperature $\mu$-metal shield inside the vacuum can of the dilution refrigerator. 

Our experimental setup enables noise power spectral density (PSD) and scalar network analyzer (SNA) measurements (see \hyperref[sec:wiring_setup]{Supplementary}). Noise calibration is performed using the Y-factor method\cite{Jebari2018ICTA, UMartel2025VBinfluence} with two $\SI{50}{\ohm}$ terminations, thermally  anchored to the mixing and still temperature stages and connected via a six-port RF switch, allowing us to convert room-temperature PSD measurements to photon flux per unit bandwidth at the switch level. This calibration is required for measuring the ICTA noise. With all switch ports open, an SNA calibration measurement is performed to calibrate the gain to the switch level, which is required for measuring the ICTA gain. The SNA measurements use the same hardware for readout as the PSD measurements, so that the attenuation of the input line to the switch is given by SNA gain divided by the gain of the output chain from the PSD measurements. This calibration is used for compression point measurements of the ICTA. Note that all calibrations are performed up to the microwave switch, subsequent ICTA measurements are, therefore, conservative estimates including losses in the package and the cable connecting to the switch.



\begin{figure*}[htbp]
\centering
\includegraphics[%
width = \textwidth]{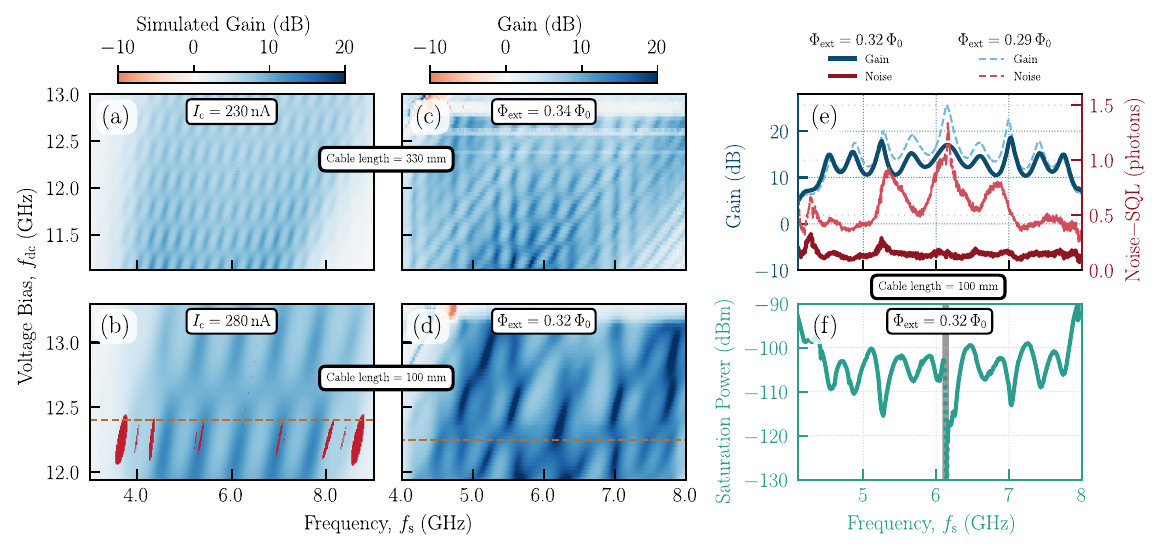}
\caption{\label{fig:fig3}
\textbf{Simulations and experimental performance of the ICTA.}
\textbf{(a)} Simulated gain of the impedance-transformed ICTA, including  realistic impedance mismatches in the cables connecting the amplifier and circulator (cable length: $\SI{330}{\milli\meter}$ @ $1/\sqrt{2}$ relative phase velocity, $\SI{55}{\ohm}$ impedance).
\textbf{(b)} Simulated gain for a $\SI{100}{\milli\meter}$ cable at $I_{\text{c}} = \SI{280}{\nano\ampere}$, near the design $f_{\text{dc}} = \SI{12}{\giga\hertz}$, used to identify voltage bias points that minimize gain ripples. The dashed light brown line at $\SI{12.6}{\giga\hertz}$ marks an optimal bias where standing waves in signal and idler modes cancel, yielding minimal ripple. Red patches indicate regions where the simulation did not converge.
\textbf{(c)} Experimental 2D gain map at relatively low $I_{\text{c}}$ ($\Phi_\text{ext} = 0.34\Phi_0$) with $\sim \SI{330}{\milli\meter}$ of transmission line between circulator and amplifier. The data shows good qualitative agreement with the simulation in (a).
\textbf{(d)} Experimental gain color plot with a $\sim \SI{100}{\milli\meter}$ transmission line at higher $I_{\text{c}}$ ($\Phi_\text{ext} = 0.32\Phi_0$) in qualitative agreement with the simulation in (b). 
\textbf{(e)} Amplifier performance at $f_{\text{dc}} = \SI{12.25}{\giga\hertz}$ (dashed line in (d)): gain (left axis) and added noise above the standard quantum limit (right axis) for the same $I_{\text{c}}$ as in panel (d) ($\Phi_\text{ext} = 0.32\Phi_0$, solid, dark) and higher $I_{\text{c}}$ ($\Phi_\text{ext} = 0.29\Phi_0$, dashed, light).
\textbf{(f)} Input-referred \SI{1}{\decibel} compression point versus frequency at same conditions as dark lines in panel (e) ($\Phi_\text{ext} = 0.32\Phi_0$). The dashed green line and greyed region near $\SI{6.125}{\giga\hertz}$ represent the degenerate bias point, where injection locking leads to high uncertainty of saturation 
power.} 

\end{figure*}

We characterize the ICTA gain as a function of voltage and flux bias using a low-amplitude probe tone to remain in the linear regime. The amplifier is voltage-biased near twice the center frequency of the design band. As shown in Fig.~\ref{fig:fig3}(c), the measured gain versus signal frequency and voltage bias at a fixed flux of \(\Phi_{\text{ext}} = 0.34\,\Phi_0\) through the SQUID exhibits the expected gain over the full designed band. Superimposed on this profile are ripples which we attribute to reflections in the approximately $\sim \SI{330}{\milli\meter}$ long transmission line from device, through the switch to the circulator. These ripples depend on both signal frequency \(f_{\text{s}}\) (vertical bands) and idler frequency \(f_{\text{dc}} - f_{\text{s}}\) (diagonal bands), and nearly vanish at certain \(f_{\text{dc}}\) due to cancellation of signal and idler modulations. We extract a ripple period of \(\SI{320}{\mega\hertz}\) from the data, in good agreement with the cable length and use it in simulations, which closely reproduce the experimental 2D gain structure (Fig.~\ref{fig:fig3}(a)).

In order to verify the origin of this ripple we have shortened the transmission line length to \(\SI{100}{\milli\meter}\) while leaving the rest of the setup unchanged. The resulting experimental gain map (Fig.~\ref{fig:fig3}(d)) exhibits a markedly larger ripple period of \(\sim\!\SI{1}{\giga\hertz}\). This period from Fig.~\ref{fig:fig3}(d) was used for the simulation shown in Fig.~\ref{fig:fig3}(b) which qualitatively agrees with the experimental results.

With the shortened-cable configuration, we optimize the Josephson frequency $f_{\text{dc}}$ to minimize gain ripple while maximizing bandwidth, after which we sweep the critical current $I_{\text{c}}$. As $I_{\text{c}}$ increases, the average gain within the target bandwidth initially rises while maintaining quantum-limited noise performance. At higher $I_{\text{c}}$, the gain continues to increase, but the noise begins to rise as well (see Fig. \ref{fig:fig3}(e)). Beyond a critical $I_{\text{c}}$, the average gain drops sharply, accompanied by a substantial noise increase, indicating the onset of parametric oscillations. Optimal performance is achieved near the target Josephson frequency and just below the onset of excess noise, at $f_{\text{dc}} = \SI{12.25}{\giga\hertz}$ and $\Phi_{\text{ext}} = 0.32\,\Phi_0$. Under these conditions, the impedance-transformed ICTA delivers $> \SI{10}{\decibel}$ gain ($\SI{13}{\decibel}$ average) over a $\SI{3.5}{\giga\hertz}$ bandwidth with an average added noise of less than 0.2 photons above the standard quantum limit (dark blue gain and dark red noise traces in Fig.~\ref{fig:fig3}(e)). At this operating point, we measure the input-referred 1 dB compression point ($P_{\text{1dB}}$) as a function of frequency [Fig.~\ref{fig:fig3}(f)], with an average value of $\SI{-106}{\dBm}$ across the bandwidth.

The experimental results, both of gain and saturation power in Fig.~\ref{fig:fig3}(e,f), are in good agreement with the performance expected from simulation (Fig.~\ref{fig:fig2}(b)) except for ripples due to standing waves in the cables. A key difference is that the degenerate amplification regime at ($f_\text{dc} = 2f_{\text{s}}$) emerges in simulations, but experimental amplification remains non-degenerate--even near half the pump frequency--due to low-frequency voltage bias noise \cite{UMartel2025VBinfluence}. Only at high signal power is the bias voltage stabilized by injection locking \cite{Lukas2021InjectionLocking} (see Fig.~\ref{fig:fig3}(f)). Our numerical simulations generally fail to converge in regimes associated with parametric oscillations or multi-stability. Experimentally, these regimes correspond to reduced gain and increased noise and the maximum gain for which simulations achieve convergence closely aligns with the highest gain attainable experimentally while maintaining noise near the quantum limit.

In a separate experiment, we have measured the emission at $f_\text{dc}$, the equivalent of the pump frequency of a JPA. We find low emission below approximately \SI{-105}{dBm} (see \hyperref[sec:power_at_pump_freq]{Supplementary}), many orders of magnitude lower than the typical pump power of a JPA and close to the input $\SI{1}{dB}$ compression point of the device, even though no particular care was taken to minimize this spurious emission. We expect that it can be significantly reduced by minimizing the impedance of the circuit seen from the junction at $f_\text{dc}$. 


In conclusion, we present the design of an impedance-transformed ICTA offering a practical, broadband, DC-powered solution for quantum-limited amplification.  Near the designed operating point, the amplifier achieves an average gain of $\SI{13}{\decibel}$ over a $\SI{3.5}{\giga\hertz}$ bandwidth, with noise below 0.2 photons above the standard quantum limit. The amplifier also exhibits a robust average saturation power of $\SI{-106}{dBm}$. While these figures of merit are well adapted to a wide range of quantum measurement tasks, such as qubit readout, experimental gain curve and saturation power are accurately predicted from simulation, allowing us to adapt the amplifier design to different needs. The practically useful figures of merit, combined with the reduced hardware complexity of a DC-powered design, promise to significantly simplify quantum-limited amplification, particularly for applications requiring multiple channels, such as quantum processor readout. The performance of the amplifier could be further improved by designing a separate idler mode to reduce frequency crowding in the signal mode, and by designing an impedance minimum at the Josephson frequency to suppress out-of-band emission. 


\begin{acknowledgments}
The authors acknowledge financial support from the Natural Sciences and Engineering Research Council of Canada (NSERC) through grants RGPIN-2025-06130 and ALLRP 565748-22 and from the Qu\'{e}bec government through Prompt Qu\'{e}bec grant 05\_AQ22.001-V3.  
\end{acknowledgments}

\bibliography{references}

@article{DSank2016MIST,
title = {Measurement-Induced State Transitions in a Superconducting Qubit: Beyond the Rotating Wave Approximation},
author = {Sank, Daniel and Chen, Zijun and Khezri, Mostafa and Kelly, J. and Barends, R. and Campbell, B. and Chen, Y. and Chiaro, B. and Dunsworth, A. and Fowler, A. and Jeffrey, E. and Lucero, E. and Megrant, A. and Mutus, J. and Neeley, M. and Neill, C. and O'Malley, P. J. J. and Quintana, C. and Roushan, P. and Vainsencher, A. and White, T. and Wenner, J. and Korotkov, Alexander N. and Martinis, John M.},
  journal = {Phys. Rev. Lett.},
  volume = {117},
  issue = {19},
  pages = {190503},
  numpages = {6},
  year = {2016},
  month = {Nov},
  publisher = {American Physical Society},
  doi = {10.1103/PhysRevLett.117.190503},
  url = {https://link.aps.org/doi/10.1103/PhysRevLett.117.190503}
}

@article{MKhezri2023MIST,
  title = {Measurement-induced state transitions in a superconducting qubit: Within the rotating-wave approximation},
  author = {Khezri, Mostafa and Opremcak, Alex and Chen, Zijun and Miao, Kevin C. and McEwen, Matt and Bengtsson, Andreas and White, Theodore and Naaman, Ofer and Sank, Daniel and Korotkov, Alexander N. and Chen, Yu and Smelyanskiy, Vadim},
  journal = {Phys. Rev. Appl.},
  volume = {20},
  issue = {5},
  pages = {054008},
  numpages = {12},
  year = {2023},
  month = {Nov},
  publisher = {American Physical Society},
  doi = {10.1103/PhysRevApplied.20.054008},
  url = {https://link.aps.org/doi/10.1103/PhysRevApplied.20.054008}
}

@incollection{GIngold1992PE,
archivePrefix = {arXiv},
arxivId = {arXiv:cond-mat/0508728v1},
author = {Ingold, Gert-Ludwig and Nazarov, Yuli V},
booktitle = {Single charge tunneling: {Coulomb} blockade phenomena in nanostructures},
chapter = {2},
doi = {10.1007/978-1-4757-2166-9_2},
editor = {Grabert, Hermann and Devoret, Michel H},
eprint = {0508728v1},
pages = {21--107},
title = {Charge Tunneling Rates in Ultrasmall Junctions},
url = {https://arxiv.org/abs/cond-mat/0508728},
series = {NATO ASI Series B},
volume = {294},
publisher = {Plenum},
address = {New York},
year = {1992}
}

@article{CBeltran2008JPA,
author={Castellanos-Beltran, M. A.
and Irwin, K. D.
and Hilton, G. C.
and Vale, L. R.
and Lehnert, K. W.},
title={Amplification and squeezing of quantum noise with a tunable Josephson metamaterial},
journal={Nature Physics},
year={2008},
month={Dec},
day={01},
volume={4},
number={12},
pages={929-931},
issn={1745-2481},
doi={10.1038/nphys1090},
url={https://doi.org/10.1038/nphys1090}
}

@Article{Bergeal2010JPA,
author={Bergeal, N.
and Schackert, F.
and Metcalfe, M.
and Vijay, R.
and Manucharyan, V. E.
and Frunzio, L.
and Prober, D. E.
and Schoelkopf, R. J.
and Girvin, S. M.
and Devoret, M. H.},
title={Phase-preserving amplification near the quantum limit with a Josephson ring modulator},
journal={Nature},
year={2010},
month={May},
day={01},
volume={465},
number={7294},
pages={64-68},
issn={1476-4687},
doi={10.1038/nature09035},
url={https://doi.org/10.1038/nature09035}
}

@article{EJeffery2014FastReadout,
  title = {Fast Accurate State Measurement with Superconducting Qubits},
  author = {Jeffrey, Evan and Sank, Daniel and Mutus, J. Y. and White, T. C. and Kelly, J. and Barends, R. and Chen, Y. and Chen, Z. and Chiaro, B. and Dunsworth, A. and Megrant, A. and O'Malley, P. J. J. and Neill, C. and Roushan, P. and Vainsencher, A. and Wenner, J. and Cleland, A. N. and Martinis, John M.},
  journal = {Phys. Rev. Lett.},
  volume = {112},
  issue = {19},
  pages = {190504},
  numpages = {5},
  year = {2014},
  month = {May},
  publisher = {American Physical Society},
  doi = {10.1103/PhysRevLett.112.190504},
  url = {https://link.aps.org/doi/10.1103/PhysRevLett.112.190504}
}

@article{Heinsoo2018MultiQreadout,
  title = {Rapid High-fidelity Multiplexed Readout of Superconducting Qubits},
  author = {Heinsoo, Johannes and Andersen, Christian Kraglund and Remm, Ants and Krinner, Sebastian and Walter, Theodore and Salath\'e, Yves and Gasparinetti, Simone and Besse, Jean-Claude and Poto\ifmmode \check{c}\else \v{c}\fi{}nik, Anton and Wallraff, Andreas and Eichler, Christopher},
  journal = {Phys. Rev. Appl.},
  volume = {10},
  issue = {3},
  pages = {034040},
  numpages = {14},
  year = {2018},
  month = {Sep},
  publisher = {American Physical Society},
  doi = {10.1103/PhysRevApplied.10.034040},
  url = {https://link.aps.org/doi/10.1103/PhysRevApplied.10.034040}
}

@article{Peter2025FastMultiplexed,
  title = {Fast Multiplexed Superconducting-Qubit Readout with Intrinsic Purcell Filtering Using a Multiconductor Transmission Line},
  author = {Spring, Peter A. and Milanovic, Luka and Sunada, Yoshiki and Wang, Shiyu and van Loo, Arjan F. and Tamate, Shuhei and Nakamura, Yasunobu},
  journal = {PRX Quantum},
  volume = {6},
  issue = {2},
  pages = {020345},
  numpages = {23},
  year = {2025},
  month = {Jun},
  publisher = {American Physical Society},
  doi = {10.1103/PRXQuantum.6.020345},
  url = {https://link.aps.org/doi/10.1103/PRXQuantum.6.020345}
}

@misc{mori2025highPowReadout,
      title={High-power readout of a transmon qubit using a nonlinear coupling}, 
      author={Cyril Mori and Vladimir Milchakov and Francesca D'Esposito and Lucas Ruela and Shelender Kumar and Vishnu Narayanan Suresh and Waël Ardati and Dorian Nicolas and Quentin Ficheux and Nicolas Roch and Tomás Ramos and Olivier Buisson},
      year={2025},
      eprint={2507.03642},
      archivePrefix={arXiv},
      primaryClass={quant-ph},
      url={https://arxiv.org/abs/2507.03642}, 
}

@Article{Jebari2018ICTA,
author={Jebari, S.
and Blanchet, F.
and Grimm, A.
and Hazra, D.
and Albert, R.
and Joyez, P.
and Vion, D.
and Est{\`e}ve, D.
and Portier, F.
and Hofheinz, M.},
title={Near-quantum-limited amplification from inelastic Cooper-pair tunnelling},
journal={Nature Electronics},
year={2018},
month={Apr},
day={01},
volume={1},
number={4},
pages={223-227},
issn={2520-1131},
doi={10.1038/s41928-018-0055-7},
url={https://doi.org/10.1038/s41928-018-0055-7}
}

@article{UMartel2025VBinfluence,
    author = {Martel, U. and Albert, R. and Blanchet, F. and Griesmar, J. and Ouellet, G. and Therrien, H. and Nehra, N. and Bourlet, N. and Peugeot, A. and Hofheinz, M.},
    title = {Influence of bias-voltage noise on the inelastic cooper-pair tunneling amplifier (ICTA)},
    journal = {Applied Physics Letters},
    volume = {126},
    number = {7},
    pages = {074001},
    year = {2025},
    month = {02},
    issn = {0003-6951},
    doi = {10.1063/5.0240842},
    url = {https://doi.org/10.1063/5.0240842},
}

@article{APaquette2022EccosorbFilt,
    author = {Paquette, Alexandre and Griesmar, Joël and Lavoie, Gabriel and Albert, Romain and Blanchet, Florian and Grimm, Alexander and Martel, Ulrich and Hofheinz, Max},
    title = {Absorptive filters for quantum circuits: Efficient fabrication and cryogenic power handling},
    journal = {Applied Physics Letters},
    volume = {121},
    number = {12},
    pages = {124001},
    year = {2022},
    month = {09},
    issn = {0003-6951},
    doi = {10.1063/5.0114887},
    url = {https://doi.org/10.1063/5.0114887},
}

@article{MutusWhite2014WideJPA,
    author = {Mutus, J. Y. and White, T. C. and Barends, R. and Chen, Yu and Chen, Z. and Chiaro, B. and Dunsworth, A. and Jeffrey, E. and Kelly, J. and Megrant, A. and Neill, C. and O'Malley, P. J. J. and Roushan, P. and Sank, D. and Vainsencher, A. and Wenner, J. and Sundqvist, K. M. and Cleland, A. N. and Martinis, John M.},
    title = {Strong environmental coupling in a Josephson parametric amplifier},
    journal = {Applied Physics Letters},
    volume = {104},
    number = {26},
    pages = {263513},
    year = {2014},
    month = {07},
    issn = {0003-6951},
    doi = {10.1063/1.4886408},
    url = {https://doi.org/10.1063/1.4886408},
}

@article{TRoy2015BeyondGBWP,
    author = {Roy, Tanay and Kundu, Suman and Chand, Madhavi and Vadiraj, A. M. and Ranadive, A. and Nehra, N. and Patankar, Meghan P. and Aumentado, J. and Clerk, A. A. and Vijay, R.},
    title = {Broadband parametric amplification with impedance engineering: Beyond the gain-bandwidth product},
    journal = {Applied Physics Letters},
    volume = {107},
    number = {26},
    pages = {262601},
    year = {2015},
    month = {12},
    issn = {0003-6951},
    doi = {10.1063/1.4939148},
    url = {https://doi.org/10.1063/1.4939148},
}

@INPROCEEDINGS{ONaman2019HighSatJPA,
  author={Naaman, O. and Ferguson, D. G. and Marakov, A. and Khalil, M. and Koehl, W. F. and Epstein, R. J.},
  booktitle={2019 IEEE MTT-S International Microwave Symposium (IMS)}, 
  title={High Saturation Power Josephson Parametric Amplifier with GHz Bandwidth}, 
  year={2019},
  volume={},
  number={},
  pages={259-262},
  doi={10.1109/MWSYM.2019.8701068}}

@article{TWhite2023HighSatJPA,
    author = {White, Theodore and others},
    title = {Readout of a quantum processor with high dynamic range Josephson parametric amplifiers},
    journal = {Applied Physics Letters},
    volume = {122},
    number = {1},
    pages = {014001},
    year = {2023},
    month = {01},
    issn = {0003-6951},
    doi = {10.1063/5.0127375},
    url = {https://doi.org/10.1063/5.0127375},
}

@article{Yaakobi2013TWPA,
  title = {Parametric amplification in Josephson junction embedded transmission lines},
  author = {Yaakobi, O. and Friedland, L. and Macklin, C. and Siddiqi, I.},
  journal = {Phys. Rev. B},
  volume = {87},
  issue = {14},
  pages = {144301},
  numpages = {9},
  year = {2013},
  month = {Apr},
  publisher = {American Physical Society},
  doi = {10.1103/PhysRevB.87.144301},
  url = {https://link.aps.org/doi/10.1103/PhysRevB.87.144301}
}

@article{Macklin2015RPM,
author = {C. Macklin  and K. O’Brien  and D. Hover  and M. E. Schwartz  and V. Bolkhovsky  and X. Zhang  and W. D. Oliver  and I. Siddiqi },
title = {A near–quantum-limited Josephson traveling-wave parametric amplifier},
journal = {Science},
volume = {350},
number = {6258},
pages = {307-310},
year = {2015},
doi = {10.1126/science.aaa8525},
URL = {https://www.science.org/doi/abs/10.1126/science.aaa8525}}

@Article{Ranadive2022KerrRev,
author={Ranadive, Arpit
and Esposito, Martina
and Planat, Luca
and Bonet, Edgar
and Naud, C{\'e}cile
and Buisson, Olivier
and Guichard, Wiebke
and Roch, Nicolas},
title={Kerr reversal in Josephson meta-material and traveling wave parametric amplification},
journal={Nature Communications},
year={2022},
month={Apr},
day={01},
volume={13},
number={1},
pages={1737},
issn={2041-1723},
doi={10.1038/s41467-022-29375-5},
url={https://doi.org/10.1038/s41467-022-29375-5}
}

@article{ONaman2022Synthesis,
  title = {Synthesis of Parametrically Coupled Networks},
  author = {Naaman, Ofer and Aumentado, Jos\'e},
  journal = {PRX Quantum},
  volume = {3},
  issue = {2},
  pages = {020201},
  numpages = {37},
  year = {2022},
  month = {May},
  publisher = {American Physical Society},
  doi = {10.1103/PRXQuantum.3.020201},
  url = {https://link.aps.org/doi/10.1103/PRXQuantum.3.020201}
}

@article{DHover2012SLUG,
    author = {Hover, D. and Chen, Y.-F. and Ribeill, G. J. and Zhu, S. and Sendelbach, S. and McDermott, R.},
    title = {Superconducting low-inductance undulatory galvanometer microwave amplifier},
    journal = {Applied Physics Letters},
    volume = {100},
    number = {6},
    pages = {063503},
    year = {2012},
    month = {02},
    issn = {0003-6951},
    doi = {10.1063/1.3682309},
    url = {https://doi.org/10.1063/1.3682309},
}

@Article{Lahteenmaki2012SJA,
author={L{\"a}hteenm{\"a}ki, Pasi
and Vesterinen, Visa
and Hassel, Juha
and Sepp{\"a}, Heikki
and Hakonen, Pertti},
title={Josephson junction microwave amplifier in self-organized noise compression mode},
journal={Scientific Reports},
year={2012},
month={Feb},
day={20},
volume={2},
number={1},
pages={276},
issn={2045-2322},
doi={10.1038/srep00276},
url={https://doi.org/10.1038/srep00276}
}

@article{MHofheinz2011BrightSide,
  title = {Bright Side of the Coulomb Blockade},
  author = {Hofheinz, M. and Portier, F. and Baudouin, Q. and Joyez, P. and Vion, D. and Bertet, P. and Roche, P. and Esteve, D.},
  journal = {Phys. Rev. Lett.},
  volume = {106},
  issue = {21},
  pages = {217005},
  numpages = {4},
  year = {2011},
  month = {May},
  publisher = {American Physical Society},
  doi = {10.1103/PhysRevLett.106.217005},
  url = {https://link.aps.org/doi/10.1103/PhysRevLett.106.217005}
}

@article{Lukas2021InjectionLocking,
  title = {Injection locking and synchronization in Josephson photonics devices},
  author = {Danner, Lukas and Padurariu, Ciprian and Ankerhold, Joachim and Kubala, Bj\"orn},
  journal = {Phys. Rev. B},
  volume = {104},
  issue = {5},
  pages = {054517},
  numpages = {14},
  year = {2021},
  month = {Aug},
  publisher = {American Physical Society},
  doi = {10.1103/PhysRevB.104.054517},
  url = {https://link.aps.org/doi/10.1103/PhysRevB.104.054517}
}

@article{RAPP_AM_AM1991,
author="RAPP, C.",
title="Effects of HPA-nonlinearity on a 4-DPSK/OFDM-signal for a digitial sound broadcasting system",
journal="Proc. Second European Conference on Satellite Communications, Liege, Belgium, Oct. 1991",
year="1991",
pages="179-184",
URL="https://cir.nii.ac.jp/crid/1571980075954482944"
}



\pagebreak
\widetext
\begin{center}
\textbf{\large Supplementary Material}
\end{center}

\setcounter{page}{1}
\setcounter{equation}{0}
\setcounter{figure}{0}
\setcounter{table}{0}
\setcounter{section}{0}
\setcounter{subsection}{0}

\renewcommand{\theequation}{S\arabic{equation}}
\renewcommand{\thefigure}{S\arabic{figure}}
\renewcommand{\thetable}{S\arabic{table}}

\renewcommand{\theHfigure}{S\arabic{figure}}
\renewcommand{\theHequation}{S\arabic{equation}} 
\renewcommand{\theHtable}{S\arabic{table}}      

\renewcommand\bibnumfmt[1]{[S#1]}
\renewcommand\citenumfont[1]{S#1}

\section{\label{sec:simulations} Simulations}

We numerically investigate gain and saturation power of the ICTA via semiclassical simulations. We separate the ICTA into a linear circuit part and the Josephson junction. The linear circuit is described by its response $a^\text{out}(\omega) = F(\omega) a^\text{in}(\omega)$ with a linear response matrix $F$ (for Frankenstein), a generalization of the scattering matrix $S$ to arbitrary boundary conditions, where each port $i$ can be a wave port (with $a^\text{in}_i$ and $a^\text{out}_i$ propagating voltage waves), a voltage-bias port (with $a^\text{in}_i$ the voltage applied to port $i$ and $a^\text{out}_i$ the current drawn by the circuit), or a current-bias port (with $a^\text{in}_i$ the current bias to port $i$ and $a^\text{out}_i$ voltage response at port $i$). We get $F = (K + L S)(M + N S)^{-1}$ with diagonal matrices $K$, $L$, $M$, $N$ with entries from table \ref{tab:fmatrixdef} and $S$ the usual voltage scattering matrix of the circuit. Note that the matrices $F$ and $K$, $L$, $M$, $N$ and the vectors $a^\text{in}$ and $a^\text{out}$ can have non-uniform units.

\begin{table}
{\arrayrulecolor{black}\setcellgapes{1ex}
\makegapedcells
\definecolor{lightgray}{gray}{0.9}
\everymath{\displaystyle}

\begin{tabular}{!{\vrule width 2\arrayrulewidth} A !{\vrule width 2\arrayrulewidth} A !{\vrule width \arrayrulewidth} A !{\vrule width \arrayrulewidth} A !{\vrule width \arrayrulewidth}  A !{\vrule width 2\arrayrulewidth}}
 \Xhline{2\arrayrulewidth}
    Type of port $i$& $K_{ii}$ & $L_{ii}$  & $M_{ii}$ & $N_{ii}$ \\ \Xhline{2\arrayrulewidth}
    voltage-bias port & $\frac{1}{Z_0}$ & $-\frac{1}{Z_0}$ & 1 & 1\\ \Xhline{\arrayrulewidth}
    current-bias port & 1  & 1  & $\frac{1}{Z_0}$ & $-\frac{1}{Z_0}$\\ \Xhline{\arrayrulewidth}
    \parbox{2.2cm}{wave port with\\impedance $Z_i$} & $\frac{1}{2}\left(1-\frac{Z_i}{Z_0}\right)$
               & $\frac{1}{2}\left(1+\frac{Z_i}{Z_0}\right)$ 
               & $\frac{1}{2}\left(1+\frac{Z_i}{Z_0}\right)$
               & $\frac{1}{2}\left(1-\frac{Z_i}{Z_0}\right)$\\ \Xhline{2\arrayrulewidth}
\end{tabular}}
\caption{\label{tab:fmatrixdef}Elements of the diagonal matrices to convert the scattering matrix $S$ of with reference impedance $Z_0$ to the Frankenstein matrix $F$.}
\end{table}

We label J the port to which the Josephson junction is connected and described it on the linear circuit side as a current bias port. The current through the Josephson junction is then the input to the linear circuit $a^\text{in}_\text{J} = I_\text{J}$ and the output of the linear circuit is the voltage applied to the Josephson junction $V_\text{J} = a^\text{out}_\text{J}$. We iteratively solve this nonlinear problem,

\begin{eqnarray}
  V_{\text{J},n}(\omega) &=& \sum_{i\neq\text{J}} F_{\mathrm{J},i}(\omega) a^\text{in}_i(\omega) + F_\mathrm{J,J} I_{\mathrm{J},n-1}(\omega)\\
  \varphi_n(t) &=& \mathcal{F}^{-1} \left\{ \frac{2eV_{\text{J},n}(\omega)}{\imath\hbar\omega} \right\} + \varphi_0 \\
  I_{\mathrm{J},n}(t) &=& I_\mathrm{C} \sin(\varphi_n(t)) \\
  I_{\mathrm{J},n}(\omega) &=& \mathcal{F}\left\{I_{\mathrm{J},n}(t)\right\},
\end{eqnarray}
starting with $I_{\mathrm{J},0}(\omega) = 0\,\, \forall\,\, \omega$. 

Here $\varphi_0$ is an integration constant defining the phase reference of the voltage bias, $\imath^2 = -1$, and $\mathcal{F}$ denotes the Fourier transform, which we approximate using a fast Fourier transform (FFT). To avoid leakage we round the voltage bias such that the Josephson frequency is a multiple of the frequency spacing of the FFT, and we minimize aliasing by zero-padding in frequency space. 

We consider the simulation converged if $|I_{\mathrm{J},n+1}(\omega) - I_{\mathrm{J},n}(\omega)| < \epsilon I_\mathrm{C}\,\, \forall\,\, \omega$ with $\epsilon = 10^{-12}$ close to the numerical precision of the calculation. We then compute the final output
\begin{equation}
  a^\text{out}_i = \sum_{j\neq \text{J}} F_{i,j} a^\text{in}_j + F_{i,\text{J}} I_\mathrm{J}.
\end{equation}

For the algorithm to converge, we need to approximate the DC voltage bias as stiff, i.e.\ $F_\text{J,DC}(0) = \SI{0}{\ohm}$, where DC denotes the DC voltage bias port to the linear circuit, so that the Josephson frequency does not depend on the DC current drawn by the junction. 

With this approximation, while convergence of the algorithm has not been fully analyzed yet, we empirically find that it is fast for low $I_\mathrm{C}$, slows down close to the threshold of parametric oscillation, and is rarely reached in presence of parametric oscillation or when multiple solutions are possible, unless initial conditions are very close to a solution. 

Quick convergence at a desired gain, therefore, gives a good indication that a design will remain stable at this gain.

\section{\label{sec:fabrication} Fabrication}
Devices are fabricated on 3-inch sapphire wafers. All lithographic patterning steps 
are performed using \SI{100}{\kilo\electronvolt} electron-beam lithography.

Fabrication begins by the depositing a \SI{100}{\nano\meter}-thick niobium 
ground plane via DC magnetron sputtering. The film is patterned using a 
\SI{300}{\nano\meter}-thick layer of CSAR~62 (AR-P~6200.13) resist and dry-etched 
in an \ce{SF6}/Ar (1:4) ICP plasma.

A \SI{200}{\nano\meter}-thick silicon nitride dielectric layer is then deposited 
by plasma-enhanced chemical vapor deposition (PECVD) using a 
\ce{SiH4}:\ce{N2} flow ratio of 1:7. For patterning this insulating 
layer, the following resist stack is employed: a bottom layer of negative-tone 
ma-N~2405 resist topped with a conductive layer of Electra~92 (AR-PC~5092) for 
charge dissipation. After electron-beam exposure and development, vias through 
the silicon nitride are etched using the same \ce{SF6}/Ar (1:4) ICP plasma recipe 
as for the niobium ground plane.

The surface is then lightly ion-milled immediately prior to sputtering a second 
\SI{100}{\nano\meter}-thick niobium layer, which forms the crossover wiring level. 
This wiring layer is patterned with positive-tone CSAR~62 resist and etched in a 
selective Ar:\ce{Cl2} (1:5) ICP plasma.

Finally, Manhattan-style aluminum Josephson junctions are fabricated using a 
standard bilayer liftoff process. The liftoff stack consists of a single layer of 
MMA~EL13 copolymer coated with two layers of PMMA (diluted in anisole). 
A conductive Electra~92 top layer is again applied during electron-beam exposure 
to prevent charging effects.
Finally, the wafer is diced into $5 \times \SI{5}{\square\milli\meter}$  chips.

\section{\label{sec:wiring_setup} Detailed Measurement Setup and Calibration Procedure}

\begin{figure}[htbp]
\centering
\includegraphics[width=0.7\textwidth]{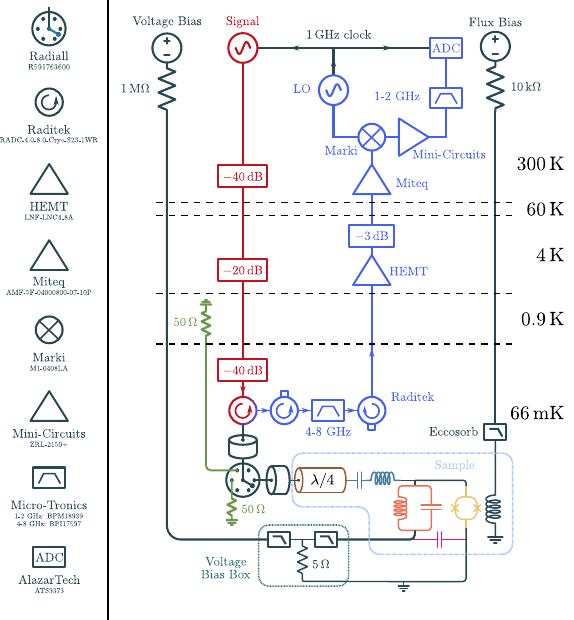} 
\caption{
\textbf{Wiring Diagram.} Schematic of the cryogenic microwave measurement setup for the impedance-transformed ICTA in the 4--8\,GHz band. The input line (red) is heavily attenuated at multiple temperature stages and routed to the device via a cryogenic circulator and a six-port switch. The output chain (blue) includes two isolators, a bandpass filter, a 4\,K HEMT amplifier, and room-temperature amplification before single-heterodyne downconversion. Two \SI{50}{\ohm} terminations anchored to the mixing chamber and still plate (light green) enable Y-factor noise calibration at the switch reference plane. DC/flux bias lines are heavily filtered at the base temperature.}

\label{fig:wiring_setup} 

\end{figure}

The impedance-transformed ICTA is characterized in the \SI{4}{}-\SI{8}{\giga\hertz} microwave frequency band using the cryogenic wiring configuration shown in Fig.~\ref{fig:wiring_setup}

\subsection{Microwave Signal Generation and Readout Chain}

Continuous-wave input signals and the local oscillator (LO) for downconversion are generated by two Rohde~\&~Schwarz SGS100A SGMA RF sources. The input signal is heavily attenuated at room temperature, the \SI{4}{\kelvin} stage, and the \SI{66}{\milli\kelvin} baseplate before reaching the device under test. It is then routed through a Raditek cryogenic circulator  and one port of a Radiall six-port cryogenic microwave switch to the input of the ICTA package.

The reflected signal from the ICTA returns through the same switch port, then passes through two additional Raditek circulators configured as isolators (to protect the device from amplifier noise) and a $\SIrange{4}{8}{\giga\hertz}$ band pass filter (Microtronics BPI17597). The signal is first amplified by a low-noise HEMT amplifier (Low Noise Factory LNC4\_8A, typical noise temperature $\sim \SI{2}{\kelvin}$) mounted at the $\SI{4}{\kelvin}$ stage, followed by room-temperature amplification. The amplified signal is downconverted in a single-heterodyne setup and digitized using the second Nyquist band of a $\SI{2}{GS/s}$ digitizer (ATS9373). Both the input RF source and the digitizer are phase-locked to the LO via a shared $\SI{1}{\giga\hertz}$ reference clock to ensure phase coherence.

This hardware configuration is used for all measurements reported in the main text: noise power spectral density (PSD), scalar network analyzer (SNA) gain and compression of the ICTA.

\subsection{\label{sec:Y-factor} Noise Calibration (Y-Factor Method)}

Calibration of the output-chain gain and system noise temperature is performed using the Y-factor technique~\cite{Jebari2018ICTA,UMartel2025VBinfluence}. Two $\SI{50}{\ohm}$ terminations are connected to the ports of the six-port cryogenic switch: one thermally anchored to the mixing chamber (baseplate, $\sim\SI{66}{\milli\kelvin}$) and one to the still plate ($\sim\SI{0.9}{\kelvin}$). These loads are thermally isolated from the switch using short ($\sim \SI{25}{mm}$) NbTi cables. By sequentially switching between these ``hot'' and ``cold'' loads while keeping all other ports open, we measure the noise power at the digitizer for two well-defined input noise temperatures. The ``cold'' temperature is not critical for this calibration as long as $k_B T_{\text{cold}} < hf/2$. The ``hot'' temperature is measured with a Ge thermometer in good thermal contact with the load. This procedure calibrates the total gain of the entire output chain (from the switch plane to the digitizer), including isolators, bandpass filter, cryogenic and room-temperature amplifiers, cables, and downconversion stage. The calibrated gain is used to refer room-temperature PSD measurements back to the input of the switch (in units of photon flux per Hz), which is essential for quantifying the ICTA-added photon noise.

\subsection{SNA and Input-Line Attenuation Calibration}

With all six ports of the switch disconnected, i.e. the common port is terminated in an open circuit, a reflection calibration of the SNA response is performed using the same readout chain. This measurement yields the combined gain of the output chain and the attenuation of the input line up to the switch plane. By dividing the magnitude of the SNA gain by the output-chain gain obtained from the Y-factor measurement, we extract the input-line attenuation (typically around \SI{100}{\decibel}) with high accuracy. This calibrated input attenuation is subsequently used to refer compression ($\SI{1}{\decibel}$) measurements to the switch plane.

Because all calibrations (Y-factor and SNA) are performed at the common reference plane defined by the microwave switch (neglecting loss in the NbTi cables between switch and ``hot''/``cold'' loads), the final ICTA characterization includes the small contributions from the cryogenic coaxial cable between the switch and the sample box.

\subsection{DC and Flux Bias Lines}

The Josephson junctions in the SQUID loop are voltage- and flux-biased using two channels of a Bilt BE2102 DC voltage source module.

\begin{itemize}
    \item \textbf{Voltage bias:} A voltage divider (\SI{5}{\ohm}\,/\,\SI{1}{\mega\ohm}) sets the voltage bias across the junctions. The bias line is heavily filtered at the base plate. The filter presents a flat \SI{5}{\ohm} output impedance up to $\sim$\SI{1}{\giga\hertz}~\cite{UMartel2025VBinfluence}.
    
    \item \textbf{Flux bias:} The flux line includes a \SI{10}{\kilo\ohm} series resistor at room temperature followed by an absorptive Eccosorb CR-124 low-pass filter anchored to the base plate with rejection $> \SI{100}{\decibel}$ in the $\SIrange{4}{8}{\giga\hertz}$ band ~\cite{APaquette2022EccosorbFilt}.
\end{itemize}

This calibration and biasing scheme ensures that gain, noise, and nonlinearity measurements of the ICTA are accurately referred to a well-defined reference plane.

\section{\label{sec:power_at_pump_freq} Emission at Josephson Frequency}

\begin{figure}[htbp]
\centering
\includegraphics[width=\textwidth]{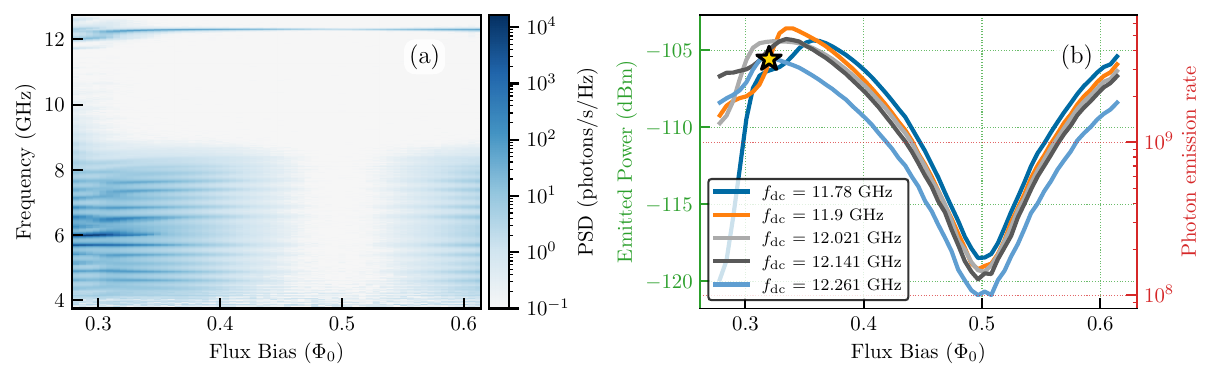} 
\caption{\label{fig:Emission_fdc} 
\textbf{Spontaneous photon emission at the Josephson frequency.}
\textbf{(a)} Power spectral density (PSD) as a function of flux-tuned critical current $I_\text{c}$ (x-axis, in units of $\Phi_0$) and emission frequency (y-axis, in GHz), measured at a fixed junction bias voltage corresponding to $f_\text{dc} = \SI{12.261}{\giga\hertz}$. Strong emission occurs exactly at $f_\text{dc}$ when the amplifier is biased for high gain.
\textbf{(b)} Photon emission power integrated from $\SIrange{11.5}{12.5}{\giga\hertz}$, for five different Josephson frequencies plotted versus $I_\text{c}$. Left axis: power referred to the switch output (dBm); right axis: equivalent photon rate (photons/s). The optimal operating point, marked by the yellow star, corresponds to $\Phi_\text{ext} \approx 0.32\,\Phi_0$}

\end{figure}
Conventional Josephson parametric amplifiers (JPAs) and traveling-wave parametric amplifiers (TWPAs) are powered from a strong pump tone (typically in the range of \SIrange{-85}{-65}{\dBm} at the device input) which must often be filtered to avoid back-action on the device under test or saturation of the subsequent measurement chain.

The ICTA operates without a microwave pump, requiring less complicated hardware to power it, but does generate radiation at $f_\text{dc} = 2eV_\text{dc}/h$ via the AC Josephson effect. Here we examine the intensity of this spontaneous photon emission. These measurements are performed in a similar setup but featuring a $\SIrange{4}{12.5}{\giga\hertz}$ measurement bandwidth and a double-heterodyne measurement setup with $\SIrange{16}{18}{\giga\hertz}$ IF1 band. 

To quantify the pump power emitted by the ICTA and enable comparison, the output noise power spectral density is recorded at room temperature and converted to an input-referred photon flux per hertz using the gain of the output chain, which is determined from Y-factor calibration (see \ref{sec:Y-factor}). Figure~\ref{fig:Emission_fdc}(a) shows the resulting emission spectrum at a fixed bias voltage of $f_\text{dc} = \SI{12.261}{\giga\hertz}$. A clear emission line appears precisely at $f_\text{dc}$, and the emission intensity increases as the flux-tuned critical current is raised to increase the gain.
The photon flux density is then integrated over a \SI{1}{\giga\hertz} bandwidth centered around $\SI{12.261}{\giga\hertz}$ for five different voltage bias points, each chosen to provide similar gain as the optimal bias point,  but with higher amplitude ripples. The integrated emitted power and corresponding photon rate are plotted in Fig.~\ref{fig:Emission_fdc}(b). Power emission stays below approximately $-\SI{105}{\dBm}$, equivalent to roughly $3.5 \times 10^{9}$ photons/s referred to the switch output. Although this emission level remains high for many practical implementations, we emphasize that it is significantly lower than typical JPA or TWPA pumps and can be easily reduced by suppressing the circuit impedance at the pump frequency in improved designs.

\section{\label{sec:saturation_power} Saturation Power}

\begin{figure}
\centering
\includegraphics[width=\textwidth]{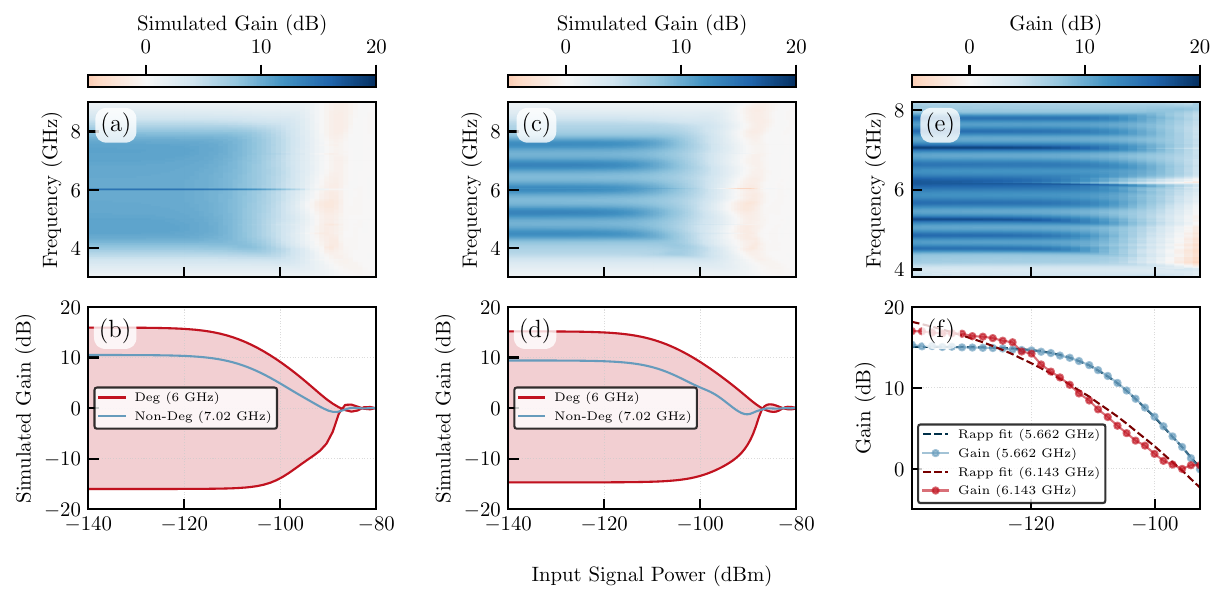} 
\caption{\label{fig:saturation_power}
\textbf{Simulations and experimental results of saturation.} \textbf{(a)} Simulated gain (color scale) as a function of input signal power and signal frequency at a fixed bias voltage corresponding to $f_\text{dc} = \SI{12}{\giga\hertz}$ and critical current $I_c = \SI{280}{\nano\ampere}$. \textbf{(b)} Amplifier gain versus input signal power for degenerate and non-degenerate parametric amplification processes. The shaded area in red corresponds to the range of degenerate gain that can be achieved by varying the phase of the signal with respect to the pump.  \textbf{(c) \& (d)} Same as (a) \& (b), with added small-magnitude gain ripples due to impedance mismatches between circulator and the device. \textbf{(e)} Experimental gain plot as a function of calibrated input signal power at the switch level on the x-axis and signal frequency on y-axis. \textbf{(f)} Measured gain as a function of signal power at degenerate (red) and non-degenerate (blue) frequency points, dashed lines are fits using the Rapp model.} 

\end{figure}
We present detailed gain compression characteristics of the amplifier as a function of input power from which the \SI{1}{\decibel} compression points in the main text are extracted. Fig.~\ref{fig:saturation_power}(a) shows the simulated gain as a function of input signal power at a fixed bias point corresponding to $f_\text{dc} = \SI{12}{\giga\hertz}$ and $I_\text{c} = \SI{280}{\nano \ampere}$, the same conditions as in Fig.~\ref{fig:fig2}(b) of the main text. The phase-locked voltage bias in the simulator enables clear separation of degenerate and non-degenerate parametric amplification modes. As a result, gain at the degenerate frequency ($f_\text{dc}/2$) is either enhanced or suppressed depending on the signal phase. 
Fig.~\ref{fig:saturation_power}(b) displays the corresponding one-dimensional gain compression curves for both regimes. To mimic experimental conditions, realistic standing waves in the line connecting the amplifier (as in Fig.~\ref{fig:fig3}(d) in the main text) are then introduced into the simulation Fig.~\ref{fig:saturation_power}(c)\&(d). Experimental results, measured at a bias of $f_\text{dc} = \SI{12.25}{\giga\hertz}$ and $\Phi_\text{ext} = 0.32\Phi_0$ (same conditions as Fig.~\ref{fig:fig3}(f)) are shown in Fig.~\ref{fig:saturation_power}(e) as a function of calibrated input power at the switch and signal frequency. In order to extract accurate \SI{1}{\decibel} compression points despite noisy data, each frequency trace is fitted using the Rapp AM-AM model:
$$P_{\text{out}} = \frac{G_0 P_{\text{in}}}{\left[1 + \left(\dfrac{G_0 P_{\text{in}}}{P_{\text{sat}}}\right)^{2p}\right]^{1/(2p)}},$$
where $G_0$ is the small-signal gain, $P_{\text{sat}}$ is the input saturation power, and $p$ is the knee smoothness parameter \cite{RAPP_AM_AM1991}. From these fits, $P_{\text{1dB}}$ is extracted and plotted versus frequency in Fig.~\ref{fig:fig3}(f). While the model fits robustly away from degeneracy, convergence fails precisely at the degenerate point—consistent with the dotted-line representation in the main text (see Fig.~\ref{fig:saturation_power}(f)). We attribute this behavior to injection locking.

\end{document}